\documentclass[superscriptaddress,
preprint,
prd,tightenlines,showpacs,nofootinbib,eqsecnum,
amsfonts,amsmath,amssymb]{revtex4}
\usepackage{bm}
\usepackage{graphicx} 
\usepackage{hyperref}
\usepackage{mathrsfs}
\usepackage{multirow}

\newcommand{\ud}{\mathrm{d}}

\usepackage{ulem}
\usepackage[usenames]{color}

\allowdisplaybreaks
\begin{document}

\title{Half-integral conservative post-Newtonian approximations\\in
  the redshift factor of black hole binaries}

\author{Luc Blanchet}\email{blanchet@iap.fr}
\affiliation{$\mathcal{G}\mathbb{R}\varepsilon{\mathbb{C}}\mathcal{O}$,
  Institut d'Astrophysique de Paris --- UMR 7095 du CNRS,
  \\ Universit\'e Pierre \& Marie Curie, 98\textsuperscript{bis}
  boulevard Arago, 75014 Paris, France}

\author{Guillaume Faye}\email{faye@iap.fr}
\affiliation{$\mathcal{G}\mathbb{R}\varepsilon{\mathbb{C}}\mathcal{O}$,
  Institut d'Astrophysique de Paris --- UMR 7095 du CNRS,
  \\ Universit\'e Pierre \& Marie Curie, 98\textsuperscript{bis}
  boulevard Arago, 75014 Paris, France}

\author{Bernard F. Whiting}\email{bernard@phys.ufl.edu}
\affiliation{Institute for Fundamental Theory, Department of Physics,
  University of Florida, Gainesville, FL 32611, USA}
\affiliation{$\mathcal{G}\mathbb{R}\varepsilon{\mathbb{C}}\mathcal{O}$,
  Institut d'Astrophysique de Paris --- UMR 7095 du CNRS,
  \\ Universit\'e Pierre \& Marie Curie, 98\textsuperscript{bis}
  boulevard Arago, 75014 Paris, France}

\date{\today}

\begin{abstract}
  Recent perturbative self-force computations (Shah, Friedman \&
  Whiting, submitted to Phys.\ Rev.~{\bf D}, arXiv:1312.1952 [gr-qc]),
  both numerical and analytical, have determined that half-integral
  post-Newtonian terms arise in the conservative dynamics of
  black-hole binaries moving on exactly circular orbits. We look at
  the possible origin of these terms within the post-Newtonian
  approximation, find that they essentially originate from non-linear
  ``tail-of-tail'' integrals and show that, as demonstrated in the
  previous paper, their first occurrence is at the 5.5PN order. The
  post-Newtonian method we use is based on a
  multipolar-post-Minkowskian treatment of the field outside a general
  matter source, which is re-expanded in the near zone and extended
  inside the source thanks to a matching argument. Applying the
  formula obtained for generic sources to compact binaries, we obtain
  the redshift factor of circular black hole binaries (without spins)
  at 5.5PN order in the extreme mass ratio limit. Our result fully
  agrees with the determination of the 5.5PN coefficient by means of
  perturbative self-force computations reported in the previously
  cited paper.
\end{abstract}


\maketitle

%%%%%%%%%%%%%%%%%%%%%%%%%%%%%%%%%%%%%

\section{Introduction}

Post-Newtonian (PN) approximations (see Ref.~\cite{Bliving13} for a
review) are well suited to describe the inspiraling phase of compact
binary systems, when the post-Newtonian parameter $\epsilon\sim v/c$
is small independently of the mass ratio $q=m_1/m_2$ between the
compact bodies. On the other hand, self-force (SF) analyses, based on
black-hole perturbation theory~\cite{MiSaTa, QuWa, DW03, GW08} (see
Refs.~\cite{PoissonLR11, Detweilerorleans, Barackorleans} for reviews),
give an accurate description of extreme mass ratio binaries for which
$q\ll 1$, even in the strong field regime. The problem of the
comparison between these two powerful methods in their common domain
of validity, that of the slow-motion and weak-field regime of an
extreme mass ratio compact binary system, has received a great deal of
attention recently~\cite{Det08, BDLW10a, BDLW10b, Shah13}.

These efforts rely on the identification of a suitable gauge invariant quantity
derived by means of the very separate and distinct SF and PN calculations. The
results can be usefully compared, regardless of their manner of computation. At
the heart of all previous comparisons lies a quantity that has come to be known as
the redshift factor or observable, and was identified and first shown by
Detweiler~\cite{Det08} to be gauge invariant for circular orbits. It can be
characterized as the redshift a photon would experience in escaping from the
small compact object to infinity along the orbital axis. It is directly
related to the particle's Killing energy that is associated with the helical
Killing symmetry. The redshift factor will be at the basis of the comparison
we pursue here.

In the most recent PN-SF comparison (see the companion
paper~\cite{Shah13}), it was found that the post-Newtonian expansion
of the redshift factor for extreme mass ratio compact binaries
contains half-integral PN terms starting at the 5.5PN order. This
result had previously been unexpected, because one may naively think
that half-integral PN terms are associated with gravitational
radiation reaction damping. However here they actually describe the
\textit{conservative} part of the dynamics, since the compact binary
moves on an exactly circular orbit, and dissipative radiation reaction
effects are explicitly neglected.

The goal of the present paper is to explain this fact using
post-Newtonian theory, and to directly compute, using PN methods, the
dominant half-integral 5.5PN coefficient for comparison with the SF
result, obtained both numerically and analytically in
Ref.~\cite{Shah13}. We shall find perfect agreement with that result,
given by Eq.~(20) in~\cite{Shah13}, showing again strong internal
consistency between analytical PN and numerical/analytical SF methods,
and their joint effectiveness in describing the dynamics of compact
binary systems.

We shall compute here the redshift factor introduced in
Ref.~\cite{Det08}, for a particle moving on an exact circular orbit
around a Schwarzschild black hole. The ensuing space-time is helically
symmetric, with a helical Killing vector $K^\alpha$ such that its
value $K_1^\alpha$ at the location of the particle (labelled 1) is
tangent to the particle's four-velocity $u_1^\alpha$, defined as usual 
with unit time-like norm. The redshift
factor $u_1^T$ is then defined geometrically by
\begin{equation}\label{uTdef}
u_1^\alpha=u_1^T K_1^\alpha \,.
\end{equation}
Adopting a coordinate system in which the helical Killing vector reads
$K^\alpha\partial_ \alpha = \partial_t + \Omega\,\partial_\varphi$ 
(which defines its normalization),
where $\Omega$ is the orbital frequency of the circular motion, the
redshift factor reduces to the $t$ component $u_1^T=u_1^t\equiv\ud
t/\ud\tau_1$ of the particle's four-velocity (where $\ud\tau_1$ is the
particle's proper time), namely
\begin{equation}\label{uT}
    u_1^T = \biggl[- g_{\alpha\beta}(y_1) \frac{v_1^\alpha
        v_1^\beta}{c^2} \biggr]^{-1/2} \,.
\end{equation}
Here $g_{\alpha\beta}(y_1)$ denotes the metric evaluated at the
particle's location $y_1^\alpha = (c t, y_1^i)$ by means of an
appropriate self-field regularization (in principle dimensional
regularization~\cite{DJSdim, BDE04}), and $v_1^\alpha\equiv\ud
y_1^\alpha/\ud t=(c,v_1^i)$ is the ordinary coordinate velocity.
%Along the way we shall shed all unnecessary ``instantaneous'' terms
%and focus primarily on the relevant ``hereditary'' contributions.

Our strategy will be to obtain first the metric $g_{\alpha\beta}$ in
the exterior of a general matter system by means of a
multipolar-post-Minkowskian expansion~\cite{BD86}, and to extend next
the validity of the solution inside the source using a matching
argument.  More precisely, we consider in a first stage a general smooth matter distribution with compact support and slow internal velocities (post-Newtonian source). The field outside the post-Newtonian source is a solution of the vacuum field equations, which is re-expanded in the exterior part of the source's near zone. The matching argument we use is based on a variant of the method of matched asymptotic expansions which has been developed to connect the exterior near-zone field to the inner field of a post-Newtonian source (see \textit{e.g.}~Ref.~\cite{Bliving13}).  At the dominant level, we will deal with a homogeneous
solution of the wave equation which, being of the type
retarded-minus-advanced, is regular all over the near zone of the
source, and thus can directly be extended by matching inside the source.  

Eventually, in a second stage, the source will be specialized to a binary point
particle system and the metric will be evaluated at the location of
one of the particles. In principle, our PN calculations are valid for
any mass ratio, but it turns out that the multipole interactions
needed at 5.5PN order are rather involved for arbitrary mass
ratios. In the extreme mass ratio (SF) limit, we shall essentially
find that only one simple multipole interaction is required, namely
the interaction between two mass monopoles and the mass quadrupole
moment (consistently with an observation made in Ref.~\cite{Shah13}),
known in the literature as a ``tail-of-tail''~\cite{B98tail}.

In the context of general relativity, tails are non-linear effects physically 
due to the backscattering  of linear waves from the space-time curvature 
generated by the total mass of the source. They are non-linear in the usual language of the PN approximation (which expands flat space-time retarded 
wave operators), since they are associated with the non-linear coupling 
between radiative multipole moments and the source's mass monopole. 
The tails imply a non-locality in time since they involve an
integral depending on the history of the source from the remote past
to the current time. They are also appropriately referred to as
``\textit{hereditary}'' contributions~\cite{BD92}, in contrast to the
``\textit{instantaneous}'' contributions which depend on the dynamics
of the source only at the current time. In this paper we shall prove
that half-integral conservative post-Newtonian terms are due to
hereditary effects. In the process, we shall shed all unnecessary
instantaneous terms and focus primarily on the relevant hereditary
contributions.

The plan of this article is as follows. In Sec.~\ref{sec:dimarg} we use
dimensionality arguments to discuss the first occurence of
half-integral conservative PN terms. In Sec.~\ref{sec:source} we
present the source terms for the so-called tail-of-tail hereditary
integrals that are responsible for the 5.5PN effect in the extreme
mass ratio limit. Sec.~\ref{sec:formula} is devoted to basic formulas
enabling us to obtain the near-zone expansion of a retarded integral,
given that of its source. Finally, in Sec.~\ref{sec:5.5PN} we obtain
the piece of the metric (in two different gauges) corresponding to the
tail-of-tail at 5.5PN order and compute the redshift factor. A most
crucial but technical proof is relegated in Appendix~\ref{app:proof}.

\section{Dimensionality arguments}
\label{sec:dimarg}

We look at the dominant occurrence of terms at \textit{half-integral}
PN orders, \textit{i.e.} at $\frac{n}{2}$PN orders where $n$ is an \textit{odd}
integer, that arise in the conservative dynamics of binary point-particles
systems, moving on exactly circular orbits. Such terms 
cannot stem from non-tail/non-hereditary sources, and may be
expected to occur first at rather high PN order. Indeed, any
instantaneous (non-tail) term at any half-integral PN order will be
zero for circular orbits, as can be shown by a simple dimensional
argument. To see this, let us look at the general structure of
instantaneous terms in the redshift factor~\eqref{uT}. We assume that
the expression of $u_1^T$ is given in the frame of the center of mass,
and has been consistently order reduced, \textit{i.e.} that all
accelerations have been replaced by the lower-order equations of
motion --- the normal practice in PN theory. We have (in order of
magnitude)
\begin{equation}\label{uTinst}
\left(u_1^T\right)_\text{inst} \sim \sum_{j, k, p, q} \,\nu^j\left(\frac{G
  m}{r_{12} c^2}\right)^k\left(\frac{\bm{v}_{12}^2}{c^2}\right)^p
\left(\frac{\bm{n}_{12}\cdot\bm{v}_{12}}{c}\right)^q\,,
\end{equation}
where $m=m_1+m_2$ is the sum of the two masses, $\nu=m_1m_2/m^2$ the
symmetric mass ratio, $r_{12}=\vert\bm{y}_1-\bm{y}_2\vert$ the
relative distance between particles and
$\bm{n}_{12}=(\bm{y}_1-\bm{y}_2)/r_{12}$ the relative
direction. Furthermore $\bm{v}_{12}^2=\dot{r}_{12}^2+r_{12}^2\Omega^2$
is the squared Euclidean norm of the relative velocity between the two
particles, and $\bm{n}_{12}\cdot\bm{v}_{12}=\dot{r}_{12}$ is the
Euclidean scalar product between the unit separation vector and the
relative velocity. In Eq.~\eqref{uTinst} we have assumed that we take
the expansion when the mass ratio $\nu\to 0$. For comparison with the
SF based calculation in linear perturbation theory, we can limit
ourselves to terms linear in $\nu$.

The simple counting of the powers of $1/c$ shows that the
post-Newtonian order of the generic term in Eq.~\eqref{uTinst} is
given by $\frac{n}{2}$PN where
\begin{equation}\label{ninst}
n=2k+2p+q-2\,.
\end{equation}
If $n$ is an odd integer, then $q$ is also an odd integer, hence
Eq.~\eqref{uTinst} contains at least one factor
$\bm{n}_{12}\cdot\bm{v}_{12}$ and vanishes for circular
orbits. The crucial point in this argument is that we are dealing with
\textit{instantaneous} (non-hereditary) terms, so that the velocity
$\bm{v}_{12}$ and unit direction $\bm{n}_{12}$ are taken at the same
time, which is the current time $t$ at which we are evaluating those
quantities. Thus there is no integration over some intermediate time
extending from the infinite past up to $t$, which would allow a
coupling between these vectors at different times.  In conclusion,
half-integral conservative post-Newtonian terms that are instantaneous
give zero, and only truly hereditary integrals can contribute.

It is known that the first hereditary integral in the near-zone metric
is the tail occurring at the 4PN order~\cite{BD88, B93}. This tail is
associated with the mass quadrupole moment, and produces both
conservative and dissipative effects. Higher-order tails are
associated with higher multipole moments (mass octupole, current
quadrupole, \textit{etc.}) and arise at higher but still integral PN
orders (5PN, 6PN, \textit{etc.}). The conservative part of these tail
effects is responsible for the appearance of logarithmic terms in the
redshift factor as well as the ADM mass and angular momentum of the
binary system, which have previously been computed at
4PN~\cite{BDLW10b, D10sf} and 5PN~\cite{BDLW10b, LBW12} orders.

Interestingly, as we shall see now, the next complicated hereditary integrals
called tails-of-tails~\cite{B98tail} do occur at half-integral PN orders, and
give a first contribution at precisely the 5.5PN order. We first simplify the
problem by noticing that, for comparison with linear SF results, any product
between two or more mass or current multipole moments $I_P$ and $J_P$ other
than the mass $M$ can be discarded, since each multipole carries in front a
mass ratio $\nu$, and we want to compute~\eqref{uT} at linear order in $\nu$.
The only moment that does not carry a factor $\nu$ is the mass monopole or
total ADM mass $M$ of the source. We thus consider only multipole interactions
of type $M \times \cdots \times M \times I_P$ or $M \times \cdots \times M
\times J_P$. 

We shall prove below that, at the dominant level, the relevant piece of the
metric is a homogeneous solution of the wave equation of the type retarded minus
advanced which is regular all over the near zone of the source. The near-zone
expansion of such a homogeneous solution, when $r=\vert\mathbf{x}\vert\to 0$,
is of the type $\hat{n}_L r^{\ell+2i}$ with $i$ being a positive integer and
$\hat{n}_L$ the symmetric trace-free (STF) angular factor. The general
structure of this term in the ``gothic'' metric deviation, corresponding
to the interaction $M \times \cdots \times M \times I_P$, is\footnote{Our
  notation is as follows: The gothic metric deviation is
  $h^{\alpha\beta}\equiv\sqrt{-g}g^{\alpha\beta}-\eta^{\alpha\beta}$ where $g$
  and $g^{\alpha\beta}$ are, respectively, the determinant and the inverse of
  $g_{\alpha\beta}$, and $\eta^{\alpha\beta}=\text{diag}(-1,1,1,1)$; $L = i_1
  \cdots i_\ell$ or $P = i_1 \cdots i_p$ denote multi-indices composed of
  $\ell$ or $p$ spatial indices (ranging from 1 to 3); $\partial_L =
  \partial_{i_1} \cdots \partial_{i_\ell}$ is the product of $\ell$
  partial derivatives $\partial_i \equiv \partial / \partial x^i$;
  $x_L = x_{i_1} \cdots x_{i_\ell}$ is the product of $\ell$ spatial
  positions $x_i$; similarly $n_L = n_{i_1} \cdots n_{i_\ell}$ is the
  product of $\ell$ unit vectors $n_i=x_i/r$; the symmetric-trace-free
  (STF) projection is indicated with a hat, i.e. $\hat{x}_L \equiv
  \text{STF}[x_L]$, $\hat{n}_L \equiv \text{STF}[n_L]$,
  $\hat{\partial}_L \equiv \text{STF}[\partial_L]$. The mass and
  current multipole moments $I_P$ and $J_P$ are STF, $I_P=\hat{I}_P$
  and $J_P=\hat{J}_P$. In the case of summed-up (dummy) multi-indices
  $L$ or $P$, we do not write the $\ell$ or $p$ summations from 1 to 3
  over the dummy indices. Symmetrization over indices is denoted by
  $T_{(ij)}=\frac{1}{2}(T_{ij}+T_{ji})$. Time-derivatives of the
  moments are indicated by superscripts $(n)$.}
\begin{equation}\label{hstruct}
h^{\alpha\beta}_{M \times \cdots \times M \times I_P} \sim
\sum_{k,p,\ell,i} \frac{G^{k}
  M^{k-1}}{c^{3k+p}}\,\hat{n}_L\left(\frac{r}{c}\right)^{\ell+2i}
\int_{-\infty}^{+\infty}\ud
u\,\kappa^{\alpha\beta}_{LP}(t,u)\,I^{(a)}_P(u)\,.
\end{equation}
Here $k$ is the number of moments in the particular interaction we are
considering; it is thus made of $k-1$ mass monopoles $M$ and one non-static
multipole $I_P$. The tensorial function $\kappa^{\alpha\beta}_{LP}(t,u)$
denotes a certain dimensionless hereditary kernel (typically a logarithmic
kernel as we shall demonstrate below). The number of time derivatives on the
moment is $a=k+p+\ell+2i+1$. Counting the powers of $1/c$ we find that the PN
order of the generic term in Eq.~\eqref{hstruct} is $n=3k+p+\ell+2i+s-2$,
where $s$ is the number of spatial indices among $\alpha\beta$, \textit{i.e.}
$s=0,1,2$ according to whether $\alpha\beta=00,0i$, or $ij$. Now we have the
inequality $\vert\ell-p\vert\leqslant s$ (``law of addition of angular
momenta'') which states that the indices on the STF tensors $\hat{n}_L$ and
$I^{(a)}_P$ must be either some free spatial indices coming from
$\alpha\beta=0i$ or $ij$, or be contracted with each others. In fact we have
$\ell=p$ when $s=0$, $\ell=p-1$ or $p+1$ when $s=1$, and $\ell=p+2$, $p$ or
$p-2$ when $s=2$. Notice that $s$ has always the same parity as $\ell -p$.
>From this we can write the PN order as
\begin{equation}\label{PNorder}
n=3k+2p+2j-2\,,
\end{equation}
where $j$ is a positive integer. For a half-integral PN order we must
have $k$ odd, hence $k=3, 5, \cdots$, since we eliminate $k=1$ which
corresponds to a linear term deprived of tail. For $k\geqslant 5$,
recalling that we have at least $p\geqslant 2$ for evolving mass
moments, we see that the PN order~\eqref{PNorder} satisfies
$n\geqslant 17$, which means at least 8.5PN order. We can thus
restrict ourselves to the case of cubic interactions $k=3$ for the
structure portrayed in Eq.~\eqref{hstruct}. In this case we have
$n=7+2p+2j$ corresponding to terms 5.5PN, 6.5PN, $\cdots$ for the mass
quadrupole $p=2$, to terms 6.5PN, 7.5PN, $\cdots$ for the mass
octupole $p=3$, and so on. The first occurrences of half-integral
orders for the current moments can be deduced from the previous
discussion by noticing that the polar tensor $\varepsilon_{ija}
J_{aP-1}$, with $\varepsilon_{ija}$ denoting the three-dimensional
Levi-Civita tensor, has the same physical dimension as $\ud I_P/\ud t$
and is endowed with one extra index.  Therefore, the relations between
$a$, $n$ and $k$, $p$, $\ell+2i$ follow formally from those obtained
for $I_P$ by making the substitutions $a\rightarrow a-1$ and
$p\rightarrow p+1$. This yields terms 6.5PN, 7.5PN, $\cdots$ for the
current quadrupole $p=2$, terms 7.5PN, 8.5PN, $\cdots$ for the current
octupole $p=3$, and so on.

We find in the end that the minimal order for which we have an
occurence of half-integral PN hereditary terms (at linear order in the
mass ratio $\nu$) is 5.5PN. It corresponds to the cubic interaction $M
\times M \times I_{ij}$, between two mass monopoles and the mass
quadrupole moment, or tail-of-tail.  Nevertheless, it should be noted
that the structure $\sim\hat{n}_L r^{\ell+2i}$ assumed in
Eq.~\eqref{hstruct} for terms of half-integral PN order is only the
starting point of a PN iteration. It should generate at higher PN
order some other terms possibly of more complicated form. The details
of this iteration depend on the adopted coordinate system. We shall
see that at 5.5PN order it plays a crucial role in harmonic
coordinates, but can be avoided by choosing appropriately another
coordinate system.

\section{Source terms for the tail-of-tail interaction}
\label{sec:source}

Tails arise from a quadratic interaction between the mass monopole
moment or ADM mass $M$ of the source, and STF non-static (propagating)
multipole moments $I_P$, $J_P$ for which $p\geqslant 2$: Dominantly
the mass quadrupole $I_{ij}$, subdominantly the mass octupole
$I_{ijk}$ and current quadrupole $J_{ij}$, and so on. If we consider
only source terms that are relevant for the dominant tail interaction,
the gothic metric deviation in harmonic coordinates (\textit{i.e.}
satisfying $\partial_\beta h^{\alpha\beta}=0$), in the vacuum region
outside the matter source, obeys
\begin{equation}\label{Boxh2}
\Box h^{\alpha\beta}_{M \times I_{ij}} = \Lambda^{\alpha\beta}_{M \times I_{ij}}
\,,
\end{equation}
where $\Box\equiv\Box_\eta$ is the flat d'Alembertian operator, and
$\Lambda_{M\times I_{ij}}$ is the gravitational source term composed
of quadratic products involving derivatives of linear terms, $h_{M}$ and
$h_{I_{ij}}$, solutions of the linearized vacuum field equations. The
metric solution of Eq.~\eqref{Boxh2} diverges at the origin $r=0$
located inside the matter source, and is supposed to be matched to the
actual post-Newtonian expansion of the field inside the source.

At cubic order the gothic metric for the tail-of-tail interaction
obeys
\begin{equation}\label{Boxh3}
\Box h^{\alpha\beta}_{M \times M \times I_{ij}} =
\Lambda^{\alpha\beta}_{M \times M \times I_{ij}} \,,
\end{equation}
where $\Lambda_{M \times M \times I_{ij}}$ is made of quadratic
products between $h_{M \times M}$ and $h_{I_{ij}}$ and between $h_{M}$
and $h_{M \times I_{ij}}$, as well as cubic products between $h_{M}$,
$h_{M}$ and $h_{I_{ij}}$. This source term has been computed in
Eqs.~(2.14)--(2.16) of Ref.~\cite{B98tail}, where it is split into a
local (instantaneous) part $\mathcal{I}_{M \times M \times I_{ij}}$
and a non-local (hereditary) part $\mathcal{H}_{M \times M \times
  I_{ij}}$:
\begin{equation}\label{calIH}
\Lambda^{\alpha\beta}_{M \times M \times I_{ij}} =
\mathcal{I}^{\alpha\beta}_{M \times M \times I_{ij}} +
\mathcal{H}^{\alpha\beta}_{M \times M \times I_{ij}} \,.
\end{equation}
Clearly the hereditary part comes from the tails that are already present in
$h_{M \times I_{ij}}$ and interact with $h_{M}$ to contribute to
the cubic source term $\Lambda_{M \times M \times I_{ij}}$. 

The solutions of Eqs.~\eqref{Boxh2} and~\eqref{Boxh3} are obtained iteratively
by applying the flat retarded integral operator, denoted
$\Box^{-1}_\text{ret}$, on the source term, but after multiplying it
by a regularization factor $r^B$ to cope with the divergence of the multipole
expansion when $r\to 0$. Analytic continuation in $B\in\mathbb{C}$ is invoked
and the finite part when $B\to 0$ provides a certain particular solution. To
ensure that the harmonic coordinate condition is satisfied at each step, one
must add to the latter solution a specific homogeneous retarded
solution~\cite{BD86}, which does not generate tail integrals when expanded in
the near zone, and can be safely ignored.

The instantaneous part of the cubic source term~\eqref{Boxh3}
explicitly reads~\cite{B98tail}\footnote{From now on we generally pose
  $G=c=1$.}
\begin{subequations}\label{sourceinst}
\begin{align}
\mathcal{I}^{00}_{M \times M \times I_{ij}} &= M^2 n_{ab} r^{-7}
\biggl\{ -516 I_{ab} - 516 r I^{(1)}_{ab} - 304 r^2 I^{(2)}_{ab}
\nonumber\\&\qquad\qquad - 76 r^3 I^{(3)}_{ab} + 108 r^4 I^{(4)}_{ab}
+ 40 r^5 I^{(5)}_{ab} \biggr\} \,,\\
%%%%%%%%%%%%%%%%%%%%%%%%%%%%%%%%%%%%%%%%%%%%%%%%%%%%%%%%%%%%%%%%
\mathcal{I}^{0i}_{M \times M \times I_{ij}} &= M^2 \hat{n}_{iab}
r^{-6} \biggl\{ 4 I^{(1)}_{ab} + 4 r I^{(2)}_{ab} - 16 r^2
I^{(3)}_{ab} + {4\over 3} r^3 I^{(4)}_{ab} - {4\over 3} r^4
I^{(5)}_{ab} \biggr\} \nonumber\\ &+ M^2 n_a r^{-6} \biggl\{
-{372\over 5} I^{(1)}_{ai} - {372\over 5} r I^{(2)}_{ai} -{232\over 5}
r^2 I^{(3)}_{ai} \nonumber\\ &\qquad\qquad- {84\over 5} r^3
I^{(4)}_{ai} + {124\over 5} r^4 I^{(5)}_{ai} \biggr\} \,, \\
%%%%%%%%%%%%%%%%%%%%%%%%%%%%%%%%%%%%%%%%%%%%%%%%%%%%%%%%%%%%%%%
\mathcal{I}^{ij}_{M \times M \times I_{ij}} &= M^2 \hat{n}_{ijab}
r^{-5} \biggl\{ -190 I^{(2)}_{ab} - 118 r I^{(3)}_{ab} - {92\over 3}
r^2 I^{(4)}_{ab} - 2 r^3 I^{(5)}_{ab} \biggr\} \nonumber\\&+
M^2\delta_{ij} n_{ab} r^{-5} \biggl\{ {160\over 7} I^{(2)}_{ab} +
{176\over 7} r I^{(3)}_{ab} - {596\over 21} r^2 I^{(4)}_{ab} -
{160\over 21} r^3 I^{(5)}_{ab} \biggr\} \nonumber\\ &+
M^2\hat{n}_{a(i} r^{-5} \biggl\{ -{312\over 7} I^{(2)}_{j)a} -
{248\over 7} r I^{(3)}_{j)a} + {400\over 7} r^2 I^{(4)}_{j)a} +
{104\over 7} r^3 I^{(5)}_{j)a} \biggr\} \nonumber\\ &+ M^2r^{-5}
\biggl\{ -12 I^{(2)}_{ij} - {196\over 15} r I^{(3)}_{ij} - {56\over 5}
r^2 I^{(4)}_{ij} - {48\over 5} r^3 I^{(5)}_{ij} \biggr\}\,.
\end{align}
\end{subequations}
Here, all the time derivatives of the quadrupole moment $I^{(p)}_{ab}$ at
evaluated at the current retarded time $t-r$ --- hence the
instantaneous character of this term. The hereditary part of the
source term~\eqref{Boxh3} is
\begin{subequations}\label{sourcetail}
\begin{align}
\mathcal{H}^{00}_{M \times M \times I_{ij}} &= M^2 n_{ab} r^{-3}
\int^{+\infty}_1 \ud x \biggl\{ 96 Q_0 I^{(4)}_{ab} + \left[ {272\over
    5} Q_1 + {168\over 5} Q_3 \right] r I^{(5)}_{ab} + 32 Q_2 r^2
I^{(6)}_{ab} \biggr\}\,,\\
%%%%%%%%%%%%%%%%%%%%%%%%%%%%%%%%%%%%%%%%%%%%%%%%%%%%%%%%%%%%%%%%%%%%
\mathcal{H}^{0i}_{M \times M \times I_{ij}} &= M^2 \hat{n}_{iab} r^{-3}
\int^{+\infty}_1 \ud x \biggl\{ - 32 Q_1 I^{(4)}_{ab} + \left[
  -{32\over 3} Q_0 + {8\over 3} Q_2 \right] r I^{(5)}_{ab} \biggr\}
\nonumber\\ &+ M^2 n_a r^{-3} \int^{+\infty}_1 \ud x \biggl\{ {96\over
  5} Q_1 I^{(4)}_{ai} + \left[ {192\over 5} Q_0 + {112\over 5} Q_2
  \right] r I^{(5)}_{ai} + 32 Q_1 r^2 I^{(6)}_{ai} \biggr\}\,,\\
%%%%%%%%%%%%%%%%%%%%%%%%%%%%%%%%%%%%%%%%%%%%%%%%%%%%%%%%%%%%%%%%%%%%
\mathcal{H}^{ij}_{M \times M \times I_{ij}} &= M^2 \hat{n}_{ijab}
r^{-3} \int^{+\infty}_1 \ud x \biggl\{ - 32 Q_2 I^{(4)}_{ab} + \left[-
  {32\over 5} Q_1 - {48\over 5} Q_3 \right] r I^{(5)}_{ab} \biggr\}
\nonumber\\ &+ M^2 \delta_{ij} n_{ab} r^{-3} \int^{+\infty}_1 \ud x
\biggl\{ - {32\over 7} Q_2 I^{(4)}_{ab} + \left[ - {208\over 7} Q_1 +
  {24\over 7} Q_3 \right] r I^{(5)}_{ab} \biggr\} \nonumber\\ &+ M^2
\hat{n}_{a(i} r^{-3} \int^{+\infty}_1 \ud x \biggl\{ {96\over 7} Q_2
I^{(4)}_{j)a} + \left[ {2112\over 35} Q_1 - {192\over 35} Q_3 \right]
r I^{(5)}_{j)a} \biggr\} \nonumber\\ &+ M^2 r^{-3} \int^{+\infty}_1
\ud x \biggl\{ {32\over 5} Q_2 I^{(4)}_{ij} + \left[ {1536\over 75}
  Q_1 - {96\over 75} Q_3 \right] r I^{(5)}_{ij} + 32 Q_0 r^2
I^{(6)}_{ij} \biggr\}\,.
\end{align}
\end{subequations}
The kernels of the above tail integrals are made of Legendre functions
of the second kind, $Q_m$, which are here computed at $x$, while the
quadrupole moments $I^{(p)}_{ab}$, all appearing inside the integrals,
are evaluated at time $t-rx$. Since $x$ ranges from 1 to $+\infty$
the hereditary character of all terms in Eq.~\eqref{sourcetail} is
evident. The Legendre function $Q_m(x)$ has a branch cut from
$-\infty$ to $1$ and is conveniently expressed in terms of the usual
Legendre polynomial $P_m(x)$ by means of the explicit formula
\begin{equation}
Q_m(x) = \frac{1}{2} P_m (x) \,\ln \left(\frac{x+1}{x-1} \right)-
\sum^{m}_{ j=1} \frac{1}{j} P_{m-j}(x) P_{j-1}(x)\,.
\end{equation}

\section{General formula for integrating the source terms}
\label{sec:formula}

For any source term of the type $\hat{n}_L S(r,t-r)$, \textit{i.e.}
which has some definite multipolarity $\ell$, and is sufficiently
regular when $r\to 0$, we can write the usual retarded integral
$\Box^{-1}_\text{ret}$ of this source as~\cite{BD86}
\begin{subequations}\label{gensol}
\begin{align}
u_L(\mathbf{x},t)\equiv \Box^{-1}_\text{ret}\Bigl[\hat{n}_L
  S(r,t-r)\Bigr] &= \int_{-\infty}^{t-r}\ud
s\,\hat{\partial}_L\left\{\frac{R\left(\frac{t-r-s}{2},s\right)
  -R\left(\frac{t+r-s}{2},s\right)}{r}\right\}\,,\label{sol}\\
%%%%%%%%%%%%%%%%%%%%%%%%%%%%%%%%%%%%%%%%%%%%%%%%%%%%%%%%%%%%%%%%%%%%%
\text{where}\quad R\left(\rho,s\right) &= \rho^\ell\int_0^\rho\ud
\lambda\,\frac{(\rho-\lambda)^\ell}{\ell!}
\left(\frac{2}{\lambda}\right)^{\ell-1}\!\!S(\lambda,s)\,.\label{R}
\end{align}
\end{subequations}
In the present case we have to apply this formula to two types of
source terms, either instantaneous or hereditary, which can
generically be written as
\begin{subequations}\label{Sinsttail}
\begin{align}
S(r,t-r) &= r^{B-k} F(t-r)\,,\label{Sinst}\\
%%%%%%%%%%%%%%%%%%%%%%%%%%%%%%%%%%%%%%%%%%%%%%%%%%%%%%%%%%%%%%%%%%%%%
\text{or}\quad S(r,t-r) &= r^{B-k}\int_1^{+\infty}\ud x\,Q_m(x)F(t-r
x)\,.\label{Stail}
\end{align}
\end{subequations}
Here $F$ represents some time derivative $I^{(p)}_{ab}$ of the quadrupole
moment. Notice the important factor $r^B$ which is systematically
included and, when $\Re(B)$ is large enough, ensures the
regularity of the source term as $r\to 0$ as well as the applicability of
the integration formula~\eqref{gensol}. Complex analytic continuation
in $B\in\mathbb{C}$ is assumed throughout.

Since, ultimately, we shall be interested in the metric at the location
of one of the particles, our goal is to compute the near-zone
expansion of the solution~\eqref{gensol} when $r\to 0$. For that
purpose it is not necessary to control the full
solution $u_L(\mathbf{x},t)$. Indeed we can obtain this expansion
directly from the near-zone expansion of the
corresponding source thanks to the following formula~\cite{B93}:
\begin{subequations}\label{formuleZP}
\begin{align}
u_L(\mathbf{x},t) &=
\hat{\partial}_L\left\{\frac{G(t-r)-G(t+r)}{r}\right\} +
\Box^{-1}_\text{inst}\bigl[\hat{n}_L \overline{S(r,t-r)}\bigr]
\,,\label{sol0}\\
%%%%%%%%%%%%%%%%%%%%%%%%%%%%%%%%%%%%%%%%%%%%%%%%%%%%%%%%%%%%%%%%%%%%%
\text{with}\quad G\left(u\right) &= \int_{-\infty}^u \ud
s\,R\left(\frac{u-s}{2},s\right)\,.\label{G}
\end{align}
\end{subequations}
The first term in Eq.~\eqref{sol0} will be of primary interest.  It is
a homogeneous solution of the wave equation which is of
retarded-minus-advanced type and is thus regular when $r\to 0$. Clearly
such a solution will be directly valid inside the matter source by virtue of a
matching argument. For later reference we note that the near-zone
expansion $r\to 0$ of this term is
\begin{equation}\label{NZexp}
\hat{\partial}_L\left\{\frac{G(t-r)-G(t+r)}{r}\right\} =
-2\hat{x}_L\sum_{k=0}^{+\infty}\frac{r^{2k}}{(2k)!!(2k+2\ell+1)!!}
G^{(2k+2\ell+1)}(t)\,.
\end{equation}
The second term in~\eqref{sol0} is a particular solution of the
inhomogeneous equation which is defined by means of the operator of
``instantaneous'' potentials as
\begin{equation}\label{Iinst}
\Box^{-1}_\text{inst}\bigl[\hat{n}_L \overline{S(r,t-r)}\bigr] =
\sum_{i=0}^{+\infty}\left(\frac{\partial}{\partial
  t}\right)^{2i}\Delta^{-1-i}\bigl[\hat{n}_L
  \overline{S(r,t-r)}\bigr]\,.
\end{equation}
Such operator acts directly on the formal
near-zone expansion of the source, indicated by the overbar, namely
\begin{equation}\label{SNZ}
\overline{S(r,t-r)} =
\sum_{j=0}^{+\infty}\frac{(-r)^j}{j!}S^{(j)}(t)\,.
\end{equation}
Note that the instantaneous operator \eqref{Iinst} is always
well-defined when acting to source terms of the type~\eqref{Sinsttail}
that are multiplied by the regularization factor $r^B$. As usual we
apply repeatedly the Poisson operators $\Delta^{-1}$ on source terms
$\sim \hat{n}_L r^{B+j}$ using analytic continuation, and consider at
the end the finite part when $B\to 0$. An important point is that the
term~\eqref{Iinst} diverges when $r\to 0$ and cannot be extended
inside the matter source. It should be matched to a full-fledge
solution of the field equations inside the source.  As we shall prove
in Appendix~\ref{app:proof}, this term will actually contribute
only at integral PN orders. Therefore the only effect at the
half-integral 5.5PN order comes from the first term in
Eq.~\eqref{sol0} containing the function $G$, which we now compute.

In order to apply the formulas~\eqref{formuleZP} explicitly we need to find
the expression of the function $G(u)$ for source terms of the
type~\eqref{Sinsttail}. This is easily done for the instantaneous source
terms~\eqref{Sinst} but is more tricky for the tail terms~\eqref{Stail}. Here
we shall give the result only for the tail terms~\eqref{Stail}. The case for
the instantaneous terms can be deduced from it by replacing the Legendre
function $Q_m(x)$ by a truncated delta-function $\delta_+(x-1)$ such that
$\int_1^{+\infty}\ud x \,\delta_+(x-1)\phi(x)=\phi(1)$, \textit{i.e.} given
formally by $\delta_+(x-1)=Y(x-1)\delta(x-1)$ where $Y$ is Heaviside's
function.

To get $G(u)$ we have to manipulate three integrations: One in the
definition of the function $G$, Eq.~\eqref{G}, one in the definition
of the function $R$, Eq.~\eqref{R}, and one present in the source term
itself, Eq.~\eqref{Stail}. These three integrations can be rearranged
after appropriate commutations of integrals, changes of variables and
integrations by parts, as
\begin{equation}\label{Gres}
G\left(u\right) = C_{k,\ell,m}(B)\int_0^{+\infty}\ud \tau\,\tau^B
F^{(k-\ell-2)}(u-\tau)\,,
\end{equation}
where the $B$-dependent coefficient is given by
\begin{equation}\label{CB}
C_{k,\ell,m}(B) =
\frac{2^\ell}{\ell!}\frac{\Gamma(B-k+\ell+3)}{\Gamma(B+1)}
\int_0^{+\infty}\ud y\,Q_m(1+y)\int_0^1\ud
z\,\frac{z^{B-k-\ell+1}(1-z)^\ell}{(2+y z)^{B-k+\ell+3}}\,,
\end{equation}
$\Gamma$ being the usual Eulerian function. Notice that, depending on
the values of $k$ and $\ell$, the function $F(u-\tau)$ in
Eq.~\eqref{Gres} will appear either with multi-time derivatives or
multi-time anti-derivatives. The formula for the
coefficient~\eqref{CB}, thanks to the use of $\Gamma$-functions, is
able to treat both cases at the same time, and is valid in either
case. Again, we have finally to take the finite part of the Laurent
expansion of the result when $B\to 0$. An alternative form of
Eq.~\eqref{CB}, in which one integration is explicitly performed,
reads
\begin{equation}\label{CBalt}
C_{k,\ell,m}(B) = \frac{\Gamma(B-k-\ell+2)}{2\Gamma(B+1)}
\sum_{i=0}^\ell\frac{(\ell+i)!}{i!(\ell-i)!}
\frac{\Gamma(B-k+\ell+3)}{\Gamma(B-k+i+3)} \int_0^{+\infty}\ud
y\,\left(\frac{y}{2}\right)^i\frac{Q_m(1+y)}{(2+y)^{B-k+2}}\,.
\end{equation}

Suppose that the coefficient~\eqref{CB} or equivalently~\eqref{CBalt}
admits the singular Laurent expansion when $B\to 0$
\begin{equation}\label{laurent}
C_{k,\ell,m}(B) = \sum_{i=-q}^{+\infty}\alpha_{(i)}\,B^i\,,
\end{equation}
with finite part coefficient $\alpha_{(0)}$, residue coefficient
$\alpha_{(-1)}$, and so on. Applying the finite part at $B=0$ we see
that the function $G(u)$ reads
\begin{equation}\label{Gtail}
G(u)=-\alpha_{(0)}F^{(k-\ell-3)}(u) +
\sum_{j=1}^q\frac{\alpha_{(-j)}}{j!}\int_0^{+\infty}\ud
\tau\,(\ln\tau)^j F^{(k-\ell-2)}(u-\tau)\,.
\end{equation}
We have performed directly the integration over $\tau$ in the first
term.  It can be checked that there are always enough time derivatives
on the quadrupole moment in $F=I^{(p)}_{ab}$ so that this term is made
of some time derivative (and not anti-derivative) of this moment. Note
also that we have discarded the contribution at $\tau=+\infty$
assuming that the quadrupole moment becomes constant in the remote
past. Finally the first term in~\eqref{Gtail} is purely instantaneous
and cannot contribute at any half-integral PN order for circular
orbits as has been shown from Eq.~\eqref{uTinst} by dimensionality
arguments.

The terms in Eq.~\eqref{Gtail} with $j\geqslant 1$ correspond to
tails. As we have seen, only tails (and tails-of-tails) can contribute
for circular orbits at the 5.5PN order. Thus, what we have to do is to
control the \textit{pole} part when $B\to 0$ of the $B$-dependent
coefficients~\eqref{CB}, and we must do that for all the source terms
in Eqs.~\eqref{sourceinst} and~\eqref{sourcetail}. We find that only
simple poles appear for all these terms at 5.5PN order, \textit{i.e.}
only the term $j=1$ in \eqref{Gtail} contributes. Hence we require
\begin{equation}\label{Gtailpole}
G^\text{tail}(u) = \alpha_{(-1)}\int_0^{+\infty}\ud
\tau\,\ln\tau\,F^{(k-\ell-2)}(u-\tau)\,.
\end{equation}

\section{Control of the 5.5PN term in the redshift factor}
\label{sec:5.5PN}

In Appendix~\ref{app:proof} we show that we do not have to consider
the second term in Eq.~\eqref{sol0}, defined by~\eqref{Iinst}, since
it contributes only at integral PN orders (4PN, 5PN, 6PN,
\textit{etc.}). This situation is fortunate: We have obtained this
term only in the form of a multipole expansion valid outside the
matter source and diverging when $r\to 0$, and to control it we would
need to invoke matching to the actual post-Newtonian field inside the
physical source.

Gathering all the results for the functions $G^\text{tail}(u)$ defined
by~\eqref{Gtailpole} for all the terms in Eqs.~\eqref{sourceinst}
and~\eqref{sourcetail}, we obtain the tail-of-tail contributions in
the gothic metric as
\begin{subequations}\label{gothicmetric}
\begin{align}
(h^{00})_{M \times M \times I_{ij}} &=
  \frac{116}{21}\frac{G^3M^2}{c^{8}}\int_0^{+\infty}\ud\,\tau\ln\tau
  \,\partial_{ab}\biggl[\frac{I_{ab}^{(3)}(t-r-\tau)-I_{ab}^{(3)}(t+r-\tau)}{r}\biggr]\,,\\ (h^{0i})_{M
    \times M \times I_{ij}} &=
  \frac{4}{105}\frac{G^3M^2}{c^{7}}\int_0^{+\infty}\ud\,\tau\ln\tau
  \,\hat{\partial}_{iab}\biggl[\frac{I_{ab}^{(2)}(t-r-\tau)-I_{ab}^{(2)}(t+r-\tau)}{r}\biggr]\nonumber\\&
  -\frac{416}{75}\frac{G^3M^2}{c^{9}}\int_0^{+\infty}\ud\,\tau\ln\tau
  \,\partial_{a}\biggl[\frac{I_{ia}^{(4)}(t-r-\tau)-I_{ia}^{(4)}(t+r-\tau)}{r}\biggr]\,,\\(h^{ij})_{M
    \times M \times I_{ij}} &=
  -\frac{32}{21}\frac{G^3M^2}{c^{8}}\int_0^{+\infty}\ud\,\tau\ln\tau
  \,\delta_{ij}\partial_{ab}\biggl[\frac{I_{ab}^{(3)}(t-r-\tau)-I_{ab}^{(3)}(t+r-\tau)}{r}\biggr]\nonumber\\&
  +\frac{104}{35}\frac{G^3M^2}{c^{8}}\int_0^{+\infty}\ud\,\tau\ln\tau
  \,\hat{\partial}_{a(i}\biggl[\frac{I_{j)a}^{(3)}(t-r-\tau)-I_{j)a}^{(3)}(t+r-\tau)}{r}\biggr]\nonumber\\&
  +\frac{76}{15}\frac{G^3M^2}{c^{10}}\int_0^{+\infty}\ud\,\tau\ln\tau
  \,\frac{I_{ij}^{(5)}(t-r-\tau)-I_{ij}^{(5)}(t+r-\tau)}{r}\,.
\end{align}
\end{subequations}
At this stage we have the important verification that the latter piece
of the metric should be separately divergence free, \textit{i.e.}
$(\partial_\beta h^{\alpha\beta})_{M \times M \times I_{ij}}=0$. This
verification is important because it tests the rather involved
formulas~\eqref{CB}--\eqref{CBalt}.

Once we have the metric~\eqref{gothicmetric} we compute its near-zone
expansion $r\to 0$ thanks to the formula~\eqref{NZexp}. We need in
fact only the leading term in that formula, corresponding to $k=0$
in~\eqref{NZexp}. In anticipation of our change from the gothic metric
$h^{\alpha\beta}$ to the usual covariant metric $g_{\alpha\beta}$, we
shall include the contribution of the spatial trace
$h^{ii}\equiv\delta_{ij}h^{ij}$ together with the $h^{00}$
component. Then we get at leading order when $r\to 0$ the expressions
\begin{subequations}\label{gothicmetricZP}
\begin{align}
(h^{00}+h^{ii})_{M \times M \times I_{ij}} &=
  -\frac{824}{1575}\frac{G^3M^2}{c^{13}}x^{ab}\int_0^{+\infty}\ud\,\tau\ln\tau
  \,I_{ab}^{(8)}(t-\tau) +
  \mathcal{O}\left(\frac{1}{c^{15}}\right)\,,\\ (h^{0i})_{M \times M
    \times I_{ij}} &=
  \frac{832}{225}\frac{G^3M^2}{c^{12}}x^{a}\int_0^{+\infty}\ud\,\tau\ln\tau
  \,I_{ia}^{(7)}(t-\tau) +
  \mathcal{O}\left(\frac{1}{c^{14}}\right)\,,\\(h^{ij})_{M \times M
    \times I_{ij}} &=
  -\frac{152}{15}\frac{G^3M^2}{c^{11}}\int_0^{+\infty}\ud\,\tau\ln\tau
  \,I_{ij}^{(6)}(t-\tau) + \mathcal{O}\left(\frac{1}{c^{13}}\right)\,.
\end{align}
\end{subequations}
The powers of $1/c$ show that this indeed corresponds to a 5.5PN
term. However, we notice that in harmonic coordinates the $ij$
component of the metric, which is of order $1/c^{11}$, can be coupled
to a Newtonian term $h^{00}=-4U_\text{ext}/c^2+\mathcal{O}(1/c^4)$,
where $U_\text{ext}$ is the Newtonian potential as seen from the
exterior of the source, to produce from the next iteration a term of
order $1/c^{13}$ comparable to that in the $00$ and $ii$ components of
the metric. The exterior Newtonian potential $U_\text{ext}$, together
with the associated ``super-potential'' $\chi_\text{ext}$ such that
$\Delta \chi_\text{ext}=2U_\text{ext}$, are defined by their multipole
expansions,
\begin{subequations}\label{Uchiext}
\begin{align}
U_\text{ext}(\mathbf{x},t) &=
G\sum_{\ell=0}^{+\infty}\frac{(-)^\ell}{\ell!}\,I_L(t)
\,\partial_L\left(\frac{1}{r}\right)\,,\\\chi_\text{ext}(\mathbf{x},t)
&= 2\Delta^{-1}U_\text{ext} =
G\sum_{\ell=0}^{+\infty}\frac{(-)^\ell}{\ell!}\,I_L(t)\,\partial_L\left(r\right)\,.
\end{align}
\end{subequations}
Thus we see that, in harmonic coordinates, we shall also have a
contribution from \textit{quartic} interactions of the type $M \times
M \times I_{ij} \times I_L$. This includes, in the particular case
$\ell=0$, the interaction $M \times M \times M \times I_{ij}$ which
can be viewed as a kind of ``tail-of-tail-of-tail''. The equation
determining this quartic interaction is readily found to be
\begin{equation}\label{eqquartic}
\Delta\Bigl[(h^{00}+h^{ii})_{M \times M \times I_{ij} \times
    I_L}\Bigr] = -\frac{608}{15}\frac{G^3
  M^2}{c^{13}}\,\partial_{ij}U_\text{ext}\int_0^{+\infty}\ud\,\tau\ln\tau
\,I_{ij}^{(6)}(t-\tau) + \mathcal{O}\left(\frac{1}{c^{15}}\right)\,,
\end{equation}
and is immediately integrated as
\begin{equation}\label{quartic}
(h^{00}+h^{ii})_{M \times M \times I_{ij} \times I_L} =
  -\frac{304}{15}\frac{G^3 M^2}{c^{13}}\,\partial_{ij}\chi_\text{ext}
  \int_0^{+\infty}\ud\,\tau\ln\tau \,I_{ij}^{(6)}(t-\tau) +
  \mathcal{O}\left(\frac{1}{c^{15}}\right)\,.
\end{equation}

Next we want to extend these results inside the matter source. This is
straightforward for the piece~\eqref{gothicmetricZP} which is regular
inside the source and is valid there as it stands. However this
requires a matching argument for the extra piece~\eqref{quartic} since
it diverges when $r\to 0$. Fortunately the problem of the matching is
easily solved by noticing that the exterior Newtonian potential
$U_\text{ext}$ and super-potential $\chi_\text{ext}$ represent the
multipole expansions of the usual Poisson potential $U$ and
super-potential $\chi$ given by
\begin{subequations}\label{Uchi}
\begin{align}
U(\mathbf{x}, t) &=
G\int\frac{\ud^3\mathbf{x}'}{\vert\mathbf{x}-\mathbf{x}'\vert}
\rho(\mathbf{x}', t)\,,\label{U}\\\chi(\mathbf{x}, t) &= 2\Delta^{-1}U
= G\int\ud^3\,\mathbf{x}'\,\vert\mathbf{x}-\mathbf{x}'\vert
\,\rho(\mathbf{x}',t)\,,\label{chi}
\end{align}
\end{subequations}
where $\rho$ is the Newtonian mass density of the source. Here we
neglect post-Newtonian corrections, and have simply used the fact that
the mass moments $I_L$ take on their usual Newtonian expressions in
the Newtonian limit. Once the metric is matched, \textit{i.e.}
$U_\text{ext}$ and $\chi_\text{ext}$ are replaced by $U$ and $\chi$,
we can consider the case where the source is a point particles binary
for which we have, at Newtonian order,
\begin{subequations}\label{Uchipp}
\begin{align}
U(\mathbf{x}, t) &= \frac{G m_1}{r_1} + \frac{G
  m_2}{r_2}\,,\label{Upp}\\\chi(\mathbf{x}, t) &= G m_1\,r_1 + G
m_2\,r_2\,.\label{chipp}
\end{align}
\end{subequations}
The metric is then complete. Coming back to the usual covariant metric
$g_{\alpha\beta}$ we find the following contributions at 5.5PN order,
\begin{subequations}\label{metric}
\begin{align}
g_{00}^\text{5.5PN} &=
\frac{412}{1575}\frac{G^3M^2}{c^{13}}x^{ab}\int_0^{+\infty}\ud\,\tau\ln\tau
\,I_{ab}^{(8)}(t-\tau) \nonumber\\ & +\frac{152}{15}\frac{G^3
  M^2}{c^{13}}\,\partial_{ab}\chi \int_0^{+\infty}\ud\,\tau\ln\tau
\,I_{ab}^{(6)}(t-\tau) +
\mathcal{O}\left(\frac{1}{c^{15}}\right)\,,\label{metric00}\\ g_{0i}^\text{5.5PN}
&=
\frac{832}{225}\frac{G^3M^2}{c^{12}}x^{a}\int_0^{+\infty}\ud\,\tau\ln\tau
\,I_{ia}^{(7)}(t-\tau) +
\mathcal{O}\left(\frac{1}{c^{14}}\right)\,,\label{metric0i}\\ g_{ij}^\text{5.5PN}
&= \frac{152}{15}\frac{G^3M^2}{c^{11}}\int_0^{+\infty}\ud\,\tau\ln\tau
\,I_{ij}^{(6)}(t-\tau) +
\mathcal{O}\left(\frac{1}{c^{13}}\right)\,.\label{metricij}
\end{align}
\end{subequations}

A priori this metric will contain both conservative and dissipative
(radiation-reaction) effects. Here we want to keep only the
conservative effects, that are compatible with the helical symmetry
and exactly circular orbits. We shall assume that the split between
conservative and dissipative effects is equivalent to a split between
``time-symmetric'' and ``time-antisymmetric'' contributions in the
following sense. Namely, we decompose the tail integrals
in~\eqref{metric} into conservative and dissipative pieces defined by
\begin{subequations}\label{consdiss}
\begin{align}
\biggl(\int_0^{+\infty}\ud\,\tau\ln\tau
\,I_{ab}^{(p)}(t-\tau)\biggr)_\text{cons} &=
\frac{1}{2}\int_0^{+\infty}\ud\,\tau\ln\tau
\,\left[I_{ab}^{(p)}(t-\tau)+I_{ab}^{(p)}(t+\tau)\right]\,,\label{cons}\\ \biggl(\int_0^{+\infty}\ud\,\tau\ln\tau
\,I_{ab}^{(p)}(t-\tau)\biggr)_\text{diss} &=
\frac{1}{2}\int_0^{+\infty}\ud\,\tau\ln\tau
\,\left[I_{ab}^{(p)}(t-\tau)-I_{ab}^{(p)}(t+\tau)\right]\,.\label{diss}
\end{align}\end{subequations}
This will be justified later when we check that the equations of
motion associated with the conservative/symmetric piece of the metric
are indeed conservative, \textit{i.e.} that the acceleration is purely
radial. Notice that there should be a logarithm $\ln r$ associated
with the conservative part of the tail integral, exactly as at 4PN and
5PN orders~\cite{BDLW10b, LBW12}. However this logarithm is an
instantaneous 5.5PN term and therefore is zero for circular orbits by
the argument~\eqref{uTinst}. Finally the conservative part of the
metric at the 5.5PN order is
\begin{subequations}\label{consmetric}
\begin{align}
\bigl(g_{00}^\text{5.5PN}\bigr)_\text{cons} &=
\frac{206}{1575}\frac{G^3M^2}{c^{13}}x^{ab}\int_0^{+\infty}\ud\,\tau\ln\tau
\,\left[I_{ab}^{(8)}(t-\tau)+I_{ab}^{(8)}(t+\tau)\right] \nonumber\\ &
+\frac{76}{15}\frac{G^3 M^2}{c^{13}}\,\partial_{ab}\chi
\int_0^{+\infty}\ud\,\tau\ln\tau
\,\left[I_{ab}^{(6)}(t-\tau)+I_{ab}^{(6)}(t+\tau)\right] +
\mathcal{O}\left(\frac{1}{c^{15}}\right)\,,\label{consmetric00}\\ \bigl(g_{0i}^\text{5.5PN}\bigr)_\text{cons}
&=
\frac{416}{225}\frac{G^3M^2}{c^{12}}x^{a}\int_0^{+\infty}\ud\,\tau\ln\tau
\,\left[I_{ia}^{(7)}(t-\tau)+I_{ia}^{(7)}(t+\tau)\right] +
\mathcal{O}\left(\frac{1}{c^{14}}\right)\,,\label{consmetric0i}\\ \bigl(g_{ij}^\text{5.5PN}\bigr)_\text{cons}
&= \frac{76}{15}\frac{G^3M^2}{c^{11}}\int_0^{+\infty}\ud\,\tau\ln\tau
\,\left[I_{ij}^{(6)}(t-\tau)+I_{ij}^{(6)}(t+\tau)\right] +
\mathcal{O}\left(\frac{1}{c^{13}}\right)\,,\label{consmetricij}
\end{align}
\end{subequations}
where we recall that the super-potential $\chi$ is given by
Eq.~\eqref{chipp}.

The metric~\eqref{consmetric} corresponds to harmonic coordinates. In
harmonic coordinates we have obtained a ``quartic'' non-linear
contribution at 5.5PN order given by the second term
in~\eqref{consmetric00}. However let us introduce new coordinates,
which have the desirable property of canceling the latter quartic
non-linear contribution, and removing the $0i$ and $ij$ components of
the metric. The coordinate transformation vector from the harmonic
coordinates to the new ones is given by
\begin{subequations}\label{etamu}
\begin{align}
\eta_{0} &=
\frac{77}{225}\frac{G^3M^2}{c^{12}}x^{ab}\int_0^{+\infty}\ud\,\tau\ln\tau
\,\left[I_{ab}^{(7)}(t-\tau)+I_{ab}^{(7)}(t+\tau)\right] +
\mathcal{O}\left(\frac{1}{c^{14}}\right)\,,\label{eta0}\\
\eta_{i} &=
-\frac{38}{15}\frac{G^3M^2}{c^{11}}x^{a}\int_0^{+\infty}\ud\,\tau\ln\tau
\,\left[I_{ia}^{(6)}(t-\tau)+I_{ia}^{(6)}(t+\tau)\right] +
\mathcal{O}\left(\frac{1}{c^{13}}\right)\,.\label{etai}
\end{align}
\end{subequations}
%
%With this coordinate transformation, the non-linear correction with
%respect to a linear gauge transformation cancels the second term
%in~\eqref{consmetric00}. We 

The coordinate transformation at the requested order, including
  the non-linear correction with respect to a linear gauge
  transformation [see \textit{e.g.} Eqs.~(6.9)--(6.10) in
  Ref.~\cite{BFeom}], reads
\begin{subequations}\label{coordtransf}
\begin{align}
\bigl({g'}_{00}^{\,\text{5.5PN}}\bigr)_\text{cons} &=
\bigl(g_{00}^{\,\text{5.5PN}}\bigr)_\text{cons} +
\frac{2}{c}\partial_t\eta_0 +
\frac{2}{c^2}\partial_i\eta_j\partial_{ij}\chi +
\mathcal{O}\left(\frac{1}{c^{15}}\right)\,,\label{transf00}\\ \bigl({g'}_{0i}^{\,\text{5.5PN}}\bigr)_\text{cons}
&= \bigl(g_{0i}^{\,\text{5.5PN}}\bigr)_\text{cons} +
\frac{1}{c}\partial_t\eta_i + \partial_i\eta_0 +
\mathcal{O}\left(\frac{1}{c^{14}}\right)\,,\label{transf0i}\\ \bigl({g'}_{ij}^{\,\text{5.5PN}}\bigr)_\text{cons}
&= \bigl(g_{ij}^{\,\text{5.5PN}}\bigr)_\text{cons} + \partial_i\eta_j
+ \partial_j\eta_i +
\mathcal{O}\left(\frac{1}{c^{13}}\right)\,.\label{transfij}
\end{align}
\end{subequations}
The non-linear term in Eq.~\eqref{transf00} cancels the second term in
Eq.~\eqref{consmetric00} and we find the simple new metric
\begin{subequations}\label{consmetricp}
\begin{align}
\bigl({g'}_{00}^{\,\text{5.5PN}}\bigr)_\text{cons} &=
\frac{428}{525}\frac{G^3M^2}{c^{13}}x^{ab}\int_0^{+\infty}\ud\,\tau\ln\tau
\,\left[I_{ab}^{(8)}(t-\tau)+I_{ab}^{(8)}(t+\tau)\right] +
\mathcal{O}\left(\frac{1}{c^{15}}\right)\,,\label{consmetricp00}\\ \bigl({g'}_{0i}^{\,\text{5.5PN}}\bigr)_\text{cons}
&=
\mathcal{O}\left(\frac{1}{c^{14}}\right)\,,\label{consmetricp0i}\\ \bigl({g'}_{ij}^{\,\text{5.5PN}}\bigr)_\text{cons}
&= \mathcal{O}\left(\frac{1}{c^{13}}\right)\,.\label{consmetricpij}
\end{align}
\end{subequations}
The computations to follow have been performed with the two
metrics~\eqref{consmetric} and~\eqref{consmetricp} giving identical
results.

Following Eq.~\eqref{uT} we compute the components of the metric
[either~\eqref{consmetric} or~\eqref{consmetricp}] at the location of
the particle 1. For this, we simply replace $x^i$ by $y_1^i$ and thus (in
a center-of-mass frame) by $X_2 x_{12}^i$ where
$x_{12}^i=y_{1}^i-y_{2}^i$ and $X_2=m_2/m$. At linear order in the
mass ratio $\nu$ we can assume that $X_2=1+\mathcal{O}(\nu)$. The term
$\partial_{ab}\chi$ in the harmonic-coordinate metric necessitates a
regularization and reads
$(\partial_{ab}\chi)_1=m_2(\delta^{ab}-n_{12}^{ab})/r_{12}$ on
particle 1.

On the other hand the quadrupole moment is given by the usual
Newtonian expression $I_{ij}=m\nu\,\hat{x}_{12}^{ij}$ and its time
derivatives are computed for circular orbits using the Newtonian
equations of motion. Similarly the ADM mass is given with this
approximation by $M=m$. Then the quadrupole moment is to be evaluated
in the past and in the future, at advanced and retarded times
$t\pm\tau$. To do that we relate the separation vector and relative
velocity at earlier and future times to the current values for
circular orbits by using
\begin{subequations}\label{xpast}
\begin{align}
x_{12}^i(t\pm\tau) &= \cos(\Omega\tau)\,x_{12}^i(t) \pm
\sin(\Omega\tau)\,v_{12}^i(t)/\Omega\,,\\ v_{12}^i(t\pm\tau) &=
\mp\Omega\,\sin(\Omega\tau)\,x_{12}^i(t) +
\cos(\Omega\tau)\,v_{12}^i(t)\,,
\end{align}
\end{subequations}
where $\Omega$ is the orbital frequancy of the circular motion. We are
then left with the integrals
\begin{subequations}\label{intlog}
  \begin{align}
     \int^{+\infty}_0 \ud\tau\,\ln\tau\, \cos(2\Omega\tau) &=
     -\frac{\pi}{4\Omega} \,, \\ \int^{+\infty}_0 \ud\tau\,\ln\tau\,
     \sin(2\Omega\tau) &= -\frac{1}{2\Omega}
     \Bigl[\ln\bigl(2\Omega\bigr) +\gamma_\text{E}\Bigr]\,.
\end{align}
\end{subequations}
We shall find that, for the conservative part of the dynamics, only
the first integral --- with the factor $\pi$ --- contributes. The
other integral (with Euler's constant $\gamma_\text{E}$) will not be
needed.

It is important also to consider the modification of the equations of
motion which is induced by the 5.5PN metric~\eqref{consmetric}. We
find that with the conservative symmetrized (half-retarded plus
half-advanced) expression~\eqref{consmetric} the modification is
purely conservative, \textit{i.e.} it only affects the relation
between the orbital frequency $\Omega$ and the coordinate separation
$r_{12}$. This is a confirmation of our
prescriptions~\eqref{consdiss}. Writing only the Newtonian and 5.5PN
terms we get
\begin{equation}\label{Omega2}
\Omega^2 = \frac{G m}{r_{12}^3}\left[1 +
  \frac{27392}{525}\nu\pi\gamma^{11/2}\right]\,,
\end{equation}
where $\gamma=G m/(r_{12}c^2)$. The inverse relation in terms of $x=(G
m\,\Omega/c^3)^{2/3}$ is
\begin{equation}\label{gam}
\gamma = x\left[1 -
  \frac{27392}{1575}\nu\pi x^{11/2}\right]\,.
\end{equation}
We have checked that the modification of the motion does not affect
the position of the center of mass so we can use the usual formulas
when going to the center-of-mass frame.

At last we have the metric on particle 1 and we insert it into
Eq.~\eqref{uT}. We then go to the frame of the center of mass and
reduce the expression to circular orbits, mindful of the
modification~\eqref{gam} to the relation between orbital separation
and frequency --- which we find does not actually contribute to the
final result. Posing then $q=m_1/m_2$ and $y=(G
m_2\Omega/c^3)^{2/3}=x(1+q)^{-2/3}$, we define the SF part to the
redshift factor as $u_1^T =
u_\text{Schw}^T(y)+q\,u_\text{SF}^T(y)+\mathcal{O}(q^2)$ and find that
the 5.5PN contribution therein is
\begin{equation}\label{uTSF}
u_\text{SF}^T = y\left[1 - \frac{13696}{525}\pi\,y^{11/2}\right] \,.
\end{equation}
We have written only the Newtonian and 5.5PN terms. This result is in
perfect agreement with the high-precision numerical and analytic
computation of the gravitational self-force reported in Eq.\ (20) of
the companion paper~\cite{Shah13}.  Analytical self-force
  calculations, essentially extending those in Refs.~\cite{Shah13,
  BiniD13} and based on the Regge-Wheeler equation, have recently 
  obtained exact results up to order 6PN~\cite{BiniD13b}, in precise 
  agreement with the high-precision results of~\cite{Shah13}. While such 
  an approach is applicable to all PN orders at first order in perturbation
  theory, our methods in principle apply to arbitrarily high order in
  the mass ratio, while also extending to higher PN order.

%\section*{Acknowledgements}

\begin{acknowledgments}
This work was partly supported by NSF Grants PHY 0855503 and PHY
1205906 to UF. BFW acknowledges sabbatical support from the CNRS
through the IAP, where part of this work was carried out.
\end{acknowledgments}

\appendix

\section{Proof that certain specific terms do not contribute at 5.5PN order}
\label{app:proof}

The second term in Eq.~\eqref{sol0} is a particular solution of the
wave equation defined by means of the operator of the instantaneous
potentials $\Box^{-1}_\text{inst}$ given in Eq.~\eqref{Iinst}. It is
crucial that such an operator acts directly on the near-zone expansion of
the source term~\eqref{SNZ}, where the source term itself is given for
this application by Eq.~\eqref{Stail} namely
\begin{equation}\label{Stail1}
S(r,t-r) = r^{B-k}\int_1^{+\infty}\ud x\,Q_m(x)F(t-r x)\,.
\end{equation}
We are looking for the hereditary tail part of the metric. Since the
operator $\Box^{-1}_\text{inst}$ is instantaneous, \textit{i.e.}  it
does not involve any integral extending over time, the only possible
tail integrals will come from the tails that are already present in
the near-zone expansion (namely $\overline{S(r,t-r)}$ when $r\to 0$)
of the source term~\eqref{Stail1}.

Let us first note that one cannot compute the near-zone expansion
of~\eqref{Stail1} by directly expanding $F(t-r x)$ under the integral
sign because the coefficients in the expansion will involve the
integral of the Legendre function $Q_m(x)$ multiplied by arbitrary
powers of $x$, which will become divergent at some stage. Hence we
split the integral~\eqref{Stail1} into a ``recent'' part from $x=1$ to
$K$, where $K$ is a constant such that $K\gg 1$, and a ``remote'' part
from $K$ up to $+\infty$. Now we are allowed to perform the Taylor
expansion of $F(t-r x)$ when $r\to 0$ into the recent part. That
expansion will be made of time derivatives $F^{(n)}(t)$ with
coefficients given by finite integrals from 1 to K of some $x^n
Q_m(x)$. Hence the expansion of the recent part is purely
instantaneous and does not contain tails. Looking for hereditary tails
we can thus concentrate our attention to the remote part of the
integral, namely
\begin{equation}\label{StailK}
\overline{S(r,t-r)}\bigg|_\text{tail} = r^{B-k}\int_K^{+\infty}\ud
x\,Q_m(x)F(t-r x)\,.
\end{equation}
In the right-hand side an overbar is implicitly understood, meaning
that the expression should be considered in the form of a near-zone
expansion. Since we assumed $K\gg 1$ we are allowed to replace the
Legendre function $Q_m(x)$ by its formal expansion when $x\to\infty$,
which is of the type $Q_m(x) \sim \sum_{p=0}^{+\infty} \,x^{-m-2p-1}$,
with some constant coefficients that we shall not need to consider
here. Thus,
\begin{equation}\label{form1}
\overline{S(r,t-r)}\bigg|_\text{tail} \sim \sum_{p=0}^{+\infty}
r^{B-k}\int_K^{+\infty}\frac{\ud x}{x^{m+2p+1}}\,F(t-r x)\,.
\end{equation}
Next we repeatedly integrate the latter integrals by parts. The
all-integrated terms will be some functions $F^{(k)}(t-K r)$ which can
be Taylor-expanded when $r\to 0$ without problem. They do not contain
tails so we ignore them. After $m+2p+1$ integrations by parts we get
\begin{equation}\label{form2}
\overline{S(r,t-r)}\bigg|_\text{tail} \sim \sum_{p=0}^{+\infty}
r^{B-k+m+2p+1}\int_K^{+\infty}\ud x\,\ln x\,F^{(m+2p+1)}(t-r x)\,.
\end{equation}
As before we do not need to write the detailed ($B$-dependent)
coefficients in front of each term. Posing next $\tau=r x$ we obtain
\begin{equation}\label{form3}
\overline{S(r,t-r)}\bigg|_\text{tail} \sim \sum_{p=0}^{+\infty}
r^{B-k+m+2p}\int_{r K}^{+\infty}\ud \tau\,\ln
\tau\,F^{(m+2p+1)}(t-\tau)\,,
\end{equation}
where again, a non-tail term (proportional to $\ln r$) has been
ignored. Finally we note that the recent part of the latter integral,
from $0$ to $r K$, can also be expanded without tails. It is then
convenient to add it back in order to complete our
result~\eqref{form3}. Thus we have identified the tail part of the
source as
\begin{equation}\label{form4}
\overline{S(r,t-r)}\bigg|_\text{tail} \sim \sum_{p=0}^{+\infty}
r^{B-k+m+2p}\int_{0}^{+\infty}\ud \tau\,\ln
\tau\,F^{(m+2p+1)}(t-\tau)\,.
\end{equation}
Following the prescription~\eqref{Iinst}, it remains to apply the
operator $\Box^{-1}_\text{inst}$. This gives
\begin{equation}\label{form5}
\Box^{-1}_\text{inst}\left[\hat{n}_L \,\overline{S(r,t-r)}
  \right]\bigg|_\text{tail} \sim
\sum_{i=0}^{+\infty}\sum_{p=0}^{+\infty} \Delta^{-i-1}\left(\hat{n}_L
r^{B-k+m+2p}\right)\int_{0}^{+\infty}\ud \tau\,\ln
\tau\,F^{(m+2p+2i+1)}(t-\tau)\,.
\end{equation}
The iterated Poisson operators $\Delta^{-i-1}$ are straightforwardly
computed and, as usual, we consider the finite part when $B\to
0$. This yields some powers of $r$ and possibly some $\ln r$ due to
poles $\sim 1/B$. Thus we get (with $a=0$ or $1$)
\begin{equation}\label{form6}
\Box^{-1}_\text{inst}\left[\hat{n}_L \,\overline{S(r,t-r)}
  \right]\bigg|_\text{tail} \!\!\!\!\sim
\sum_{i=0}^{+\infty}\sum_{p=0}^{+\infty}\hat{n}_L r^{-k+m+2p+2i+2}(\ln
r)^a\int_{0}^{+\infty}\ud \tau\,\ln \tau\,F^{(m+2p+2i+1)}(t-\tau)\,.
\end{equation}
If we restore all the powers of $c$'s and $G$'s together with the fact
that $F$ is composed of a mass squared $M^2$ times a time derivative
of a quadrupole moment $I_{ab}$, we end up with
\begin{equation}\label{form7}
\Box^{-1}_\text{inst}\left[\hat{n}_L \,\overline{S(r,t-r)}
  \right]\bigg|_\text{tail} \!\!\!\!\sim G^3 M^2 \sum_{i, p}\frac{\hat{n}_L
  r^{-k+m+2p+2i+2}(\ln r)^a}{c^{13-k+m+2p+2i}}\!\int_{0}^{+\infty}\ud
\tau\,\ln \tau\,I_{ab}^{(8-k+m+2p+2i)}(t-\tau)\,.
\end{equation}

Let us look at the actual source for that particular interaction
$M^2\times I_{ab}$ as given by Eqs.~\eqref{sourcetail} --- as
explained already, we can ignore the non-tail part~\eqref{sourceinst}
of the source. We observe that, for all the terms in
Eqs.~\eqref{sourcetail}, the combination $k+m+\ell$ is always an
\textit{odd} integer. Furthermore, using the fact that the space
indices among $\alpha\beta=00,0i,ij$ must be distributed between the
indices of $\hat{n}_L$ and $I_{ab}^{(p)}$, we see that $\ell$ must be
even in the $00$ and $ij$ components of the metric and odd in the $0i$
components, see also the discussion before Eq.~\eqref{PNorder}. We
thus conclude from Eq.~\eqref{form7} that the powers of $1/c$ are even
in the $00$ and $ij$ components and odd in the $0i$ components, which
means precisely that all the terms in Eq.~\eqref{form7} have
necessarily integral PN orders. Closer inspection of~\eqref{form7}
with the explicit values of $k$, $m$ and $\ell$ in the
source~\eqref{sourcetail} shows that these terms are necessarily of
order 4PN, 5PN, 6PN and so on, but can never arise at 5.5PN order.

In conclusion, we have proved that only the first term in
Eq.~\eqref{sol0} contributes at the 5.5PN order, and this is what we
have computed in the text. A related issue is that the second term in
Eq.~\eqref{sol0}, that we have investigated in this Appendix, is in
fact divergent when $r\to 0$ --- indeed see
\textit{e.g.}~\eqref{form7} which involves negative powers of $r$,
when $k=+3$ say, which is a typical term in the
source~\eqref{sourcetail}. Thus the second term in~\eqref{sol0} cannot
be continued inside the source by itself. It has to be matched to the
actual PN expansion of the field inside the source. Only the first
term in Eq.~\eqref{sol0}, which is a regular homogeneous solution of
the wave equation, is valid inside the source and can be continued
there. This is why we could compute it at the location of one of the
particles in a binary system. The other term, by contrast,
necessitates a matching procedure which we do not control in the
present work. However, past experience with tails (\textit{e.g.} in
Ref.~\cite{BDLW10b}) indicates that one does not need a complete
matching in order to compute the tails inside the source, essentially
because they contribute to the radiation reaction and can be
determined as ``boundary conditions'' set outside the
source. Therefore we do not expect that the second term in
Eq.~\eqref{sol0} should contribute to the present tail-of-tail
effect. Regardless, in this Appendix we have directly proven that the
PN order of such a term is necessarily integral and cannot be 5.5PN.

\bibliography{BFW_final_Mar2014}

\end{document}